\documentclass[aps,pra,reprint,superscriptaddress]{revtex4-2}

\usepackage{graphicx}
\usepackage{dcolumn}
\usepackage{bm}

\usepackage{xcolor}
\usepackage{hyperref}
\hypersetup{colorlinks=true}
\usepackage[braket, qm]{qcircuit}

\begin{document}

\title{Para-particle oscillator simulations on a trapped ion quantum computer}

\author{C. Huerta Alderete}
\email[e-mail: ]{aldehuer@gmail.com}
\affiliation{Joint Quantum Institute, Department of Physics, University of Maryland, College Park, MD 20742, USA}
\affiliation{Information Sciences, Los Alamos National Laboratory, Los Alamos, NM 87545, USA}
\affiliation{Materials Physics and Applications Division, Los Alamos National Laboratory, Los Alamos, NM 87545, USA.
}
\author{Alaina M. Green}
\affiliation{Joint Quantum Institute, Department of Physics, University of Maryland, College Park, MD 20742, USA}
\author{Nhung H. Nguyen}
\affiliation{Joint Quantum Institute, Department of Physics, University of Maryland, College Park, MD 20742, USA}
\author{Yingyue Zhu}
\affiliation{Joint Quantum Institute, Department of Physics, University of Maryland, College Park, MD 20742, USA}
\author{Norbert M. Linke}
\affiliation{Joint Quantum Institute, Department of Physics, University of Maryland, College Park, MD 20742, USA}
\affiliation{Duke Quantum Center and Department of Physics, Duke University, Durham, North Carolina 27708, USA}
\author{B. M. Rodr\'iguez-Lara}
\affiliation{Tecnologico de Monterrey, Escuela de Ingenier\'ia y Ciencias, Ave. Eugenio Garza Sada 2501, Monterrey, N. L., Mexico, 64849}

\date{\today}
\begin{abstract}
    Deformed oscillators allow for a generalization of the standard fermions and bosons, namely, for the description of para-particles. Such particles, while indiscernible in nature, can represent good candidates for descriptions of physical phenomena like topological phases of matter.  Here, we report the digital quantum simulation of para-particle oscillators by mapping para-particle states to the state of a qubit register, which allow us to identify the para-particle oscillator Hamiltonian as an $XY$ model, and further digitize the system onto a universal set of gates. In both instances, the gate depth grows polynomially with the number of qubits used. To establish the validity of our results, we experimentally simulate the dynamics of para-fermions and para-bosons, demonstrating full control of para-particle oscillators on a quantum computer. 
    Furthermore, we compare the overall performance of the digital simulation of dynamics of the driven para-Fermi oscillator to a recent analog quantum simulation result. 
\end{abstract}

\maketitle

\section{Introduction}
Deformations of the harmonic oscillator have been investigated since their introduction in the 1950s \cite{Green1953p270,Greenberg1965pB1155,Wigner1950p711,Yang1951p788,Mahdifar2008p063814,Uhdre2022p013703,Plyushchay1997p619}. Amongst the most interesting representations is the parity-deformed algebra \cite{Calogero1969p2191,Vasiliev1991p1115,Plyushchay1997p619}, mainly because it allows for the description of the so-called para-particles, a generalization of the fermions and bosons \cite{Green1953p270,Greenberg1965pB1155}. Similarly to the usual particles, para-particles are split into two families, para-fermions and para-bosons, and are further classified by a parameter of deformation, or order of para-quantization, $p\geq 1$.
Standard fermions and bosons are para-particles of order $p=1$.
Para-particles are distinguished by their spin and Hilbert space dimensionality; para-fermions have half-integer spin and a finite-dimensional representation, while para-bosons have integer spin and an infinite-dimensional representation. 
These alternative particles, while theoretically well defined, are unlikely to be realized in nature \cite{Greenberg1965pB1155}. Still, para-particles have applications in a variety of different areas, from dark matter/dark energy \cite{Ebadi2013p057,Nelson2016p034039,Kitabayashi2018p043504}, the physics of solid excitations \cite{Safonov1991p109}, statistical thermodynamics \cite{Hama1992p149,Stoilva2020p126421}, characterization of non-classical properties of light \cite{HuertaAlderete2017p043835,Wei_Min2001p283, Mojaveri2018p346,Mojaveri2018p529,Mojaveri2018p1850134,Dehghani2019p1950104} as well as possible applications in optics \cite{RodriguezWalton2020p043840,HurtadoMolina2021p412698,Kockum2017p2045,Cai2020pnwaa196,Mojaveri2020227,Mojaveri2021p115102,Stoilova2019p135201} and generalized Boson sampling \cite{Kuo2022p}.
Para-particle oscillators are described by effective creation and annihilation operators with (anti-) commutation relations,
\begin{eqnarray} \label{eq:pF_comm}
    \left[ \hat{A}_{pF}, \hat{A}_{pF}^{\dagger} \right] &=& 2 \left(  \frac{p}{2} -\hat{\mathcal{N}}_{pF} \right)\hat{\mathcal{R}}_{pF}, \\ \label{eq:pB_comm}
    \left[ \hat{A}_{pB}, \hat{A}^{\dagger}_{pB}\right] &=& 1 + (p-1) \hat{\mathcal{R}}_{pB}, \\
    \left\{ \hat{\mathcal{R}}_{\vartheta}, \hat{A}_{\vartheta} \right\} &=&\left\{ \hat{\mathcal{R}}_{\vartheta}, \hat{A}^{\dagger}_{\vartheta} \right\} = 0, \qquad \hat{\mathcal{R}}^2_{\vartheta} = 1. \nonumber
\end{eqnarray}
Here, $\hat{A}_{\vartheta} \left(\hat{A}^{\dagger}_{\vartheta}\right)$ is the annihilation (creation) operator, $\hat{\mathcal{N}}_{\vartheta}$ the number operator and $\hat{\mathcal{R}}_{\vartheta}=e^{-i\pi \hat{\mathcal{N}}_{\vartheta}}$ the parity operator of para-fermions, $\vartheta=pF$, and para-bosons, $\vartheta=pB$.
The parity-deformed algebra allows for the description of para-fermions of even order, and para-bosons of any order.
This representation returns standard bosons when $p=1$ but does not provide standard fermions. 

To enable empirical studies of particles obeying para-statistics, some methods for simulating and characterizing para-particles have been proposed \cite{HuertaAlderete2016p414001,HuertaAlderete2017p013820,HuertaAlderete2018p11572,HurtadoMolina2021p412698}.
Recently, the dynamics of para-particle oscillators were simulated by the authors using a trapped ion experiment in which a spin-$1/2$ system was coupled to two motional modes simultaneously \cite{HuertaAlderete2021p}.
Here, as an alternative to the analog simulation, we use a digital quantum circuit to simulate the dynamics of driven para-particle oscillators.
The observable of interest is the expectation value of the para-particle number operator, $\langle \mathcal{N}_{\vartheta}\rangle$. This quantity allows for a direct comparison between the analog and digital simulations of a para-Fermi oscillator of order $p=2$. Furthermore, we use it to characterize a transition in the para-particle number statistics of a driven para-Bose oscillator.

\section{Qubit mapping and measurements}
\begin{figure}[h]
	\includegraphics[scale=1]{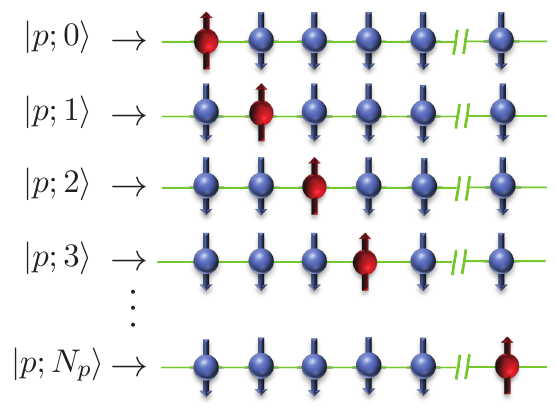}
	\caption{Representation of para-particle Fock states, $\vert p; n \rangle$, in a chain on $N_{p}+1$ qubits where the $n$th qubit is flipped.}
	\label{fig:fig1}
\end{figure}

A general para-particle system can be written as a linear combination of product states.
These can be mapped onto qubit states and prepared by flipping the relevant qubits. 
Inspired by the scheme introduced by Somma {\em et. al.} \cite{Somma2003p189}, we  propose a direct mapping between the states of the two systems. To this end, we limit the number $N_p$ of para-particles in our system, which can be represented in an array of $N_{p}+1$ qubits, see Fig. \ref{fig:fig1}.
Further, we use the actions of the parity-deformed ladder operators, Eqs. (\ref{eq:pF_comm})-(\ref{eq:pB_comm}), on para-particle Fock states, to map the Hamiltonian of a driven para-particle oscillator to qubits.

\subsection{Para-fermions}
A para-Fermi oscillator of order $p=2 N_{p}$ can be characterized by the commutation relations given in Eq. (\ref{eq:pF_comm}).
The actions of the para-Fermi operators on the para-Fermi state $\vert p; n \rangle$ with $n=0, \dots, 2N_{p}$ are given by,
\begin{eqnarray}\label{eq:pf_actions}
\hat{A}^{\dagger}_{pF} \vert p; n \rangle &=& \phi(p,n+1) \vert p; n + 1 \rangle, \nonumber \\
\hat{A}_{pF} \vert p; n \rangle &=& \phi(p,n) \vert p; n - 1 \rangle, \nonumber \\
\hat{N}_{pF} \vert p; n \rangle &=& n \vert p; n \rangle,
\end{eqnarray}
with\begin{equation}\label{eq:pf_weight}
    \phi (p,n) = \sqrt{\frac{1}{2}( p + 1)  + \frac{1}{2}\left(n-p-1\right) (-1)^{p+n}}.
\end{equation}
Recently, it was shown that the evolution of the lowest energy state under a driven para-Fermi oscillator dynamics resembles a binomial state \cite{HuertaAlderete2018p11572}.
Therefore, it produces single-frequency oscillations of the excitation number for small para-particle order, and a multiple-frequency beat as the order increases.
Hence, we are interested in the evolution of the lowest-energy para-Fermi state of even order, $\vert p;0 \rangle$, under the Hamiltonian,
\begin{eqnarray} \label{eq:HpF}
\hat{H} &=& g \left(\hat{A}_{pF} + \hat{A}_{pF}^{\dagger}\right),
\end{eqnarray}
that couples neighboring para-Fermi states with coupling strength $g$. The number of qubits needed to simulate a para-Fermi oscillator of order $p=2N_{p}$  is $2N_{p}+1$. The following operator mapping fulfills
Eqs. (\ref{eq:pf_actions}):
\begin{eqnarray}\label{eq:map2}
\hat{A}_{pF}^{\dagger} &=& \sum_{m=0}^{2N_{p}-1} \frac{1}{4}\phi(p,m+1) \hat{\sigma}_{-}^{m}\hat{\sigma}_{+}^{m+1}, \nonumber \\
\hat{\mathcal{N}}_{pF} &=& \sum_{m=0}^{2N_{p}} \frac{m}{2} \left(\hat{\sigma}_{0}^{m} + \hat{\sigma}_{z}^{m}\right),
\end{eqnarray}
where $\hat{\sigma}_{0}$ denotes the $2\times2$ Identity matrix, $\hat{\sigma}_{j}$ with $j=x,y,z$ the Pauli matrices and $\hat{\sigma}_{\pm}=\hat{\sigma}_{x}\pm \hat{\sigma}_{y}$.
In this representation, the Hamiltonian of interest, Eq. (\ref{eq:HpF}), takes the form of an $XY$ model,
\begin{eqnarray}
\hat{H} = g \sum_{m=0}^{2N_{p}-1}\frac{\phi(p,m+1)}{8} \left(\hat{\sigma}_{x}^{m} \hat{\sigma}_{x}^{m+1} +\hat{\sigma}_{y}^{m} \hat{\sigma}_{y}^{m+1}\right).
\end{eqnarray}
This model is the qubit representation of a para-Fermi oscillator, parametrized by the order $p$ and the coupling strength $g$, that can be easily implemented on a quantum compiler. 

The evolution of the lowest energy state under the Hamiltonian given in Eq. (\ref{eq:HpF}) corresponds to a vacuum displaced para-Fermi state;
\begin{equation}
    \vert \phi \rangle = \hat{D}_{pF}(\alpha) \vert p; 0 \rangle,
\end{equation}
with $\hat{D}_{pF}(\alpha) = e^{i\alpha \left( \hat{A}^{\dagger}_{pF} + \hat{A}_{pF}\right)}$ and $\alpha= gt$. 
To characterize those states, we use the expectation value of the para-Fermi number operator, $\langle \mathcal{N}_{pF} \rangle$. This measurement can be directly compared with the recent analog simulation realized in a trapped-ion experiment \cite{HuertaAlderete2021p}. In the qubit representation,
\begin{eqnarray}\nonumber
\langle \hat{\mathcal{N}}_{pF} \rangle &=& \sum_{m=0}^{2 N_{p}} \frac{m}{2} \left(\langle \hat{\sigma}_{z}^{m} \rangle + 1 \right),
\end{eqnarray} 
which requires only measurements of the vacuum displaced para-Fermi state in the $z$-basis.

\subsection{Para-bosons}
A para-Bose oscillator of order $p$ can be characterized by the commutation relation given in Eq. (\ref{eq:pB_comm}).
The actions of $\hat{A}^{\dagger}_{pB}$ and $\hat{A}_{pB}$ on the para-Bose Fock state $\vert p; n \rangle$ are given by
\begin{eqnarray}\label{eq:actions}
&&\hat{A}^{\dagger}_{pB} \vert p; 2 n \rangle = \sqrt{2n+p} \vert p; 2 n + 1 \rangle, \nonumber \\
&&\hat{A}^{\dagger}_{pB} \vert p; 2 n + 1\rangle = \sqrt{2n+2} \vert p; 2 n + 2 \rangle, \nonumber \\
&&\hat{A}_{pB} \vert p; 2 n \rangle = \sqrt{2n} \vert p; 2 n - 1 \rangle, \nonumber \\
&&\hat{A}_{pB} \vert p; 2 n + 1 \rangle = \sqrt{2n+p} \vert p; 2 n \rangle .
\end{eqnarray}
To map the infinite-dimensional representation of para-Bose oscillators onto qubits we limit the number of basis states $\vert p; n \rangle$ to $N_{p}$  where $N_{p}$ is the maximum number of para-bosons.
The matrix representation is now $(N_{p}+1)\times(N_{p}+1)$ dimensional.
As we have a finite subspace, the para-Bose commutation relation, Eq. (\ref{eq:pB_comm}), changes with $N_{p}$ as:
\begin{eqnarray}
\left[\hat{A}_{pB}, \hat{A}^{\dagger}_{pB} \right] = 1 + (p-1)\hat{\mathcal{R}}_{pB} - \beta~  \hat{A}^{\dagger~N_{p}}_{pB} \hat{A}^{N_{p}}_{pB},
\end{eqnarray}
with \begin{eqnarray}
\beta =\left\{ \begin{array}{cc}
\frac{N_{p}+1}{(N_{p}-1)!!} \frac{(p-2)!!}{(N_{p}+p-1)!!} & N_{p}-~\mathrm{odd}, \\
\frac{N_{p}+p}{N_{p}!!} \frac{(p-2)!!}{(N_{p}+p-2)!!} & N_{p}-~\mathrm{even}.
\end{array}\right.
\end{eqnarray}
Here, the notation $(x)!!$ is used to indicate a double factorial. As one would expect, $\beta \rightarrow 0$ when $N_{p} \rightarrow \infty$. 
This means that the cut-off imposed by finite $N_{p}$ induces an error which reduces with increasing qubit number.
We can now associate $N_{p} + 1$ qubits per state as shown in Fig. \ref{fig:fig1}.
Similarly to the para-Fermi case, we are interested in the evolution of the vacuum state under the Hamiltonian,
\begin{equation}\label{eq:HpB}
    \hat{H} = g \left( \hat{A}_{pB} + \hat{A}_{pB}^{\dagger}\right),
\end{equation}
that couples neighboring para-Bose states with coupling strength $g$. By using the definitions in Eq. (\ref{eq:actions}), we map the para-Bose operators to spin-$1/2$ operators,
\begin{eqnarray} \label{eq:map}
\hat{A}^{\dagger}_{pB} &=& \sum_{m=0}^{N_{p}}\zeta(p,m)~ \hat{\sigma}_{-}^{m}\hat{\sigma}_{+}^{m+1}, \nonumber \\ 
\hat{\mathcal{N}}_{pB}  &=& \sum_{m=0}^{N_{p}} \frac{m}{2}\left(\hat{\sigma}_{0}^{m} +  \hat{\sigma}_{z}^{m} \right).
\end{eqnarray}
The weight or coupling function, $\zeta(p,m)$, in the ladder operator, changes for each pair of qubits $(m,m+1)$,
\begin{eqnarray}\label{eq:pb_weight}
\zeta(p,m) = \sqrt{m +\frac{1}{2}(1+p)-\frac{(-1)^{m}}{2}(1-p)}.
\end{eqnarray}
If $p=1$, we recover the mapping proposed for standard bosons \cite{Somma2003p189}. As in the para-Fermi case, the Hamiltonian of interest, Eq. (\ref{eq:HpB}), takes the form of an $XY$ model,
\begin{eqnarray}
\hat{H} = g \sum_{m=0}^{N_{p}} \frac{1}{2}\zeta(p,m) \left(\hat{\sigma}_{x}^{m} \hat{\sigma}_{x}^{m+1} +\hat{\sigma}_{y}^{m} \hat{\sigma}_{y}^{m+1}\right).
\end{eqnarray}

It is worth mentioning that this representation is, in essence, the same as the introduced for para-Fermi operators, up to a weight function that depends on the para-particle order. While both para-particle oscillators can be described as an $XY$ model, the qubit representation of a para-Bose oscillator corresponds to a truncated Hilbert space. 

The evolution of the vacuum state under the Hamiltonian given in Eq. (\ref{eq:HpB}) corresponds to a vacuum displaced para-Bose state,
\begin{equation}
    \vert \phi \rangle = \hat{D}_{pF}(\alpha) \vert p; 0 \rangle,
\end{equation}
with $\hat{D}_{pB}(\alpha) = e^{i\alpha \left( \hat{A}^{\dagger}_{pB} + \hat{A}_{pB}\right)}$ and $\alpha= gt$, and will be used to characterize a unique feature of para-Bose oscillators, this is, that its statistics can be controlled via the para-particle order, $p$. To see this, we look at the Mandel $Q$ parameter \cite{Mandel1979p205}, \begin{eqnarray}\label{eq:MandelQ}
Q = \frac{\langle \hat{\mathcal{N}}_{pB}^2 \rangle - \langle \hat{\mathcal{N}}_{pB} \rangle^2}{\langle \hat{\mathcal{N}}_{pB} \rangle} -1,
\end{eqnarray} 
which reveals the type of statistics in the system: Poissonian ($Q=0$), super-Poissonian ($Q>0$), or sub-Poissonian ($Q<0$); and investigate (numerically) the error imposed by the cut-off $N_{p}$, for representations with increasing dimension and compare it to the analytical solution in the para-Bose frame \cite{HuertaAlderete2017p043835}.
In the qubit representation, the mean para-Bose photon-number and its variance are,
\begin{eqnarray} \label{eq:meas} \nonumber
\langle \hat{\mathcal{N}}_{pB} \rangle &=& \sum_{m=0}^{N_{p}} \frac{m}{2} \left(\langle \hat{\sigma}_{z}^{m} \rangle + 1 \right), \\
\langle \hat{\mathcal{N}}_{pB}^2 \rangle &=& \sum_{m=0}^{N_{p}} \frac{m^{2}}{2} \left( \langle \hat{\sigma}_{z}^{m} \rangle + 1 \right),
\end{eqnarray} 
then, determining $Q$ requires only measurements of the vacuum-displaced para-Bose state in the $z$-basis.
\begin{figure}[t!]
	\includegraphics[scale=1]{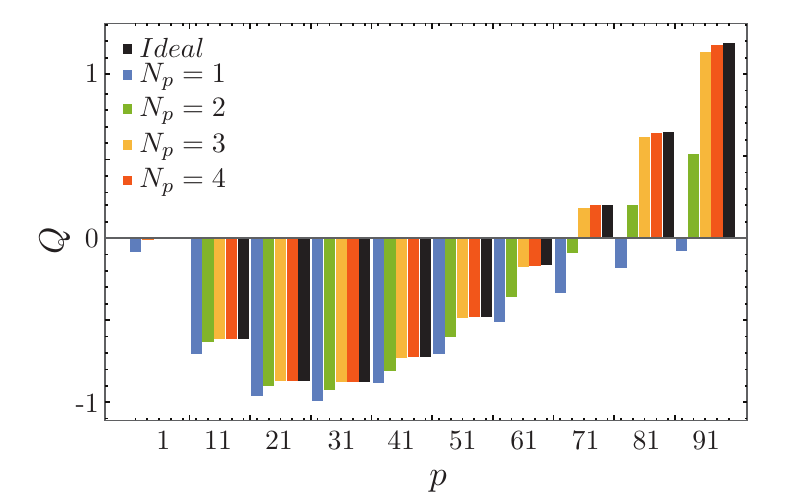}
	\caption{Theoretical study of the effect of the cut-off and finite size for the para-Boson system. Comparison of the Mandel $Q$ parameter for different values of $N_p$ where a para-Bose state is prepared with complex parameter $\alpha=0.3$ and para-particle order $p$.}
	\label{fig:Fig2}
\end{figure}
By using Eq. (\ref{eq:MandelQ}), in Fig. \ref{fig:Fig2} we show the behavior of the Mandel Q parameter as a function of $p$ and $N_{p}$ with parameter $\alpha = 0.3$. For this particular value of $\alpha$ we need at least $N_{p}=3$ to be close to the ideal values (black bars).

\section{Hardware and digitization}
\begin{figure}[t]
	\centering
	\includegraphics[scale=1]{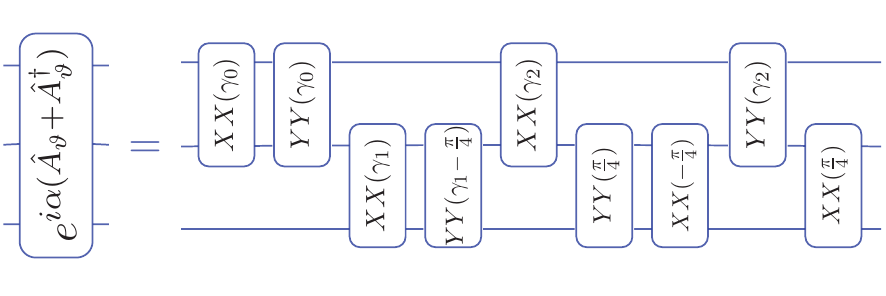}
	\caption{Circuit decomposition for a para-particle displacement operator in a three--qubit representation.}
	\label{fig:circuit}
\end{figure}

To realize a digital para-particle simulation we turn to our experimental setup, which constitutes a programmable quantum computer.
It consists of a chain of seven individual $^{171}$Yb$^{+}$ ions confined in a linear Paul trap and laser-cooled close to their motional ground state \cite{Debnath2016p63}.
Five of these are used to encode qubits in the form of a pair of states in the hyperfine-split $^{2}S_{1/2}$ ground level with an energy difference of $12.642821$ GHz, which is insensitive to magnetic field fluctuations to first order.
The qubits are initialized to $\vert 0 \rangle$ by optical pumping and read out by state-dependent fluorescence detection \cite{Olmschenk2007p052314}.
Our universal gate set consists of single-qubit rotations, or $R$ gates, and two-qubit entangling interactions, or $XX$ gates \cite{Debnath2016p63}, which are driven by applying two counter-propagating Raman beams derived from a 355-nm mode-locked laser. 
After executing the quantum circuit, the state populations are corrected for independently measured state preparation and measurement (SPAM) errors \cite{Shen2012p053053}.
For more details on the experimental setup, see, for example, Ref. \cite{Debnath2016p63}. 
To break down our simulation into these building blocks, we factorize the para-particle displacement operator, $\hat{D}_{\vartheta} $, in the qubit representation.
Because the qubit representations (\ref{eq:map2}) and (\ref{eq:map}) are the same up to a weight factor, the circuit representation is the same.
We chose a three-qubit system to simulate para-particles since the truncation error is sufficiently low, see Fig. (\ref{fig:Fig2}).
Then, according to the representations (\ref{eq:map2}) and (\ref{eq:map}) the displacement operators can be written as,
\begin{eqnarray}
e^{i\alpha \left( \hat{A}^{\dagger}_{\vartheta} + \hat{A}_{\vartheta}\right)} &=& \exp\left[i \frac{c_{\vartheta}(0)}{2} \left( \hat{\sigma}_{x}^{0} \hat{\sigma}_{x}^{1}  + \hat{\sigma}_{y}^{0} \hat{\sigma}_{y}^{1} \right)\right. \nonumber \\
&& \quad \left. +  i\frac{c_{\vartheta}(1)}{2} \left( \hat{\sigma}_{x}^{1} \hat{\sigma}_{x}^{2} + \hat{\sigma}_{y}^{1} \hat{\sigma}_{y}^{2} \right)\right],\nonumber 
\end{eqnarray}
with the weight function,
\begin{eqnarray} \nonumber
    c_{\vartheta}(m) &=& \left\{ \begin{array}{ll}
        \alpha \phi(p, m), & \text{for para-fermions},\\
         \alpha \zeta(p,m), & \text{for para-bosons}, 
    \end{array} \right.
\end{eqnarray}
where $\zeta$ and $\phi$ are given in Eqs. (\ref{eq:pf_weight}) and (\ref{eq:pb_weight}). 
From the interaction between consecutive neighbors, we define $\hat{u}_{n} = \hat{\sigma}_{x}^{n} \hat{\sigma}_{x}^{n+1} + \hat{\sigma}_{y}^{n} \hat{\sigma}_{y}^{n+1}$ and for the interaction where three qubits are involved, $\hat{v}_{n}=  \hat{\sigma}_{x}^{n} \hat{\sigma}_{z}^{n+1} \hat{\sigma}_{y}^{n+2}- \hat{\sigma}_{y}^{n} \hat{\sigma}_{z}^{n+1} \hat{\sigma}_{x}^{n+2}$. We can identify the following commutation relations,
\begin{eqnarray}\nonumber
& \left[\hat{u}_{n}, \hat{u}_{n+1} \right] = 2i \hat{v}_{n}, \quad \left[\hat{u}_{n+1}, \hat{v}_{n} \right] = 2i \hat{u}_{n}, \quad \left[\hat{v}_{n}, \hat{u}_{n} \right] = 2i \hat{u}_{n+1} & \\ \nonumber
& \left[\hat{u}_{n},\left[\hat{u}_{n+1}, \hat{v}_{n}\right]\right] + \left[\hat{u}_{n+1},\left[\hat{v}_{n}, \hat{u}_{n}\right]\right] + \left[\hat{v}_{n},\left[\hat{u}_{n}, \hat{u}_{n+1}\right]\right] = 0.& \nonumber\\
\end{eqnarray}
This means that for any three consecutive neighbors we can close an algebra related to $su(2)$ \cite{Ban1993p1347},
and as a consequence, factorize the three-qubit displacement operator as:
\begin{eqnarray}
e^{i\alpha \left( \hat{A}^{\dagger}_{\vartheta} + \hat{A}_{\vartheta}\right)}&=& e^{i \gamma_{0} \hat{u}_{0}} e^{i \gamma_{1} \hat{u}_{1}} e^{i \gamma_{2} \hat{v}_{0}},
\end{eqnarray}
where the weight factors, $\gamma_{j}$, contain information about the total qubit number $\hat{\mathcal{N}}_{\vartheta}$, the para-particle order $p$, the coupling strength $g$, and the time evolution $t$.
Sometimes it is easy to find an analytical solution for the $\gamma_{j}$ via well-known methods, such as the Baker-Campbell-Hausdorff formulas \cite{Ban1993p1347}.
Even when this is not the case, a numerical approximation for $\gamma_{j}$ is sufficient to achieve the results presented here.
As we are factorizing the exponential, our description provides an exact solution of the para-particle propagator and does not require approximate evolution methods like Trotterization.
From the construction of the operators $\hat{u}_{n}$ and $\hat{v}$ notice that, $ \left[ \hat{\sigma}_{x}^{n} \hat{\sigma}_{x}^{n+1}, \hat{\sigma}_{y}^{n} \hat{\sigma}_{y}^{n+1}\right] = \left[\hat{\sigma}_{x}^{n} \hat{\sigma}_{z}^{n+1} \hat{\sigma}_{y}^{n+2} , \hat{\sigma}_{y}^{n} \hat{\sigma}_{z}^{n+1} \hat{\sigma}_{x}^{n+2} \right] = 0 $, then 
\begin{eqnarray}
e^{i\alpha \left( \hat{A}^{\dagger}_{\vartheta} + \hat{A}_{\vartheta}\right)} &=& e^{i \gamma_{0} \hat{\sigma}_{x}^{0}\hat{\sigma}_{x}^{1}} e^{i \gamma_{0} \hat{\sigma}_{y}^{0}\hat{\sigma}_{y}^{1}} e^{i \gamma_{1} \hat{\sigma}_{x}^{1}\hat{\sigma}_{x}^{2}} e^{i \gamma_{1} \hat{\sigma}_{y}^{1}\hat{\sigma}_{y}^{2}} \times \nonumber \\
&& \times e^{i \gamma_{2} \hat{\sigma}_{x}^{0} \hat{\sigma}_{z}^{1}\hat{\sigma}_{y}^{2}} e^{i \gamma_{2} \hat{\sigma}_{y}^{0}\hat{\sigma}_{z}^{1}\hat{\sigma}_{y}^{3}}.
\label{eq:displacement}
\end{eqnarray}
By using this decomposition for the three--qubit representation of the displacement operator, the circuit shown in Fig. \ref{fig:circuit} can be generated, and implemented on the quantum computer.
This decomposition, is not unique but was chosen in a way that favors gate cancellation in the displacement operator.
Qubit states that do not belong to the subspace of para-particle Fock-states are only populated due to gate errors.
As a result, a post-selection scheme can be applied to the probabilistic outcome of the circuit.
Any runs of the circuit resulting in measurements of states out of the para-particle Hilbert space are discarded.

\section{Results and discussion}
We present the experimental results for a para-Fermi oscillator of order $p=2$ on a three--qubit system and coupling strength $g=0.02$, and compare with its analog simulation reported in Ref. \cite{HuertaAlderete2021p}, where the interaction of a singe spin-$1/2$ and two orthogonal motional modes is mapped to a driven para-Fermi oscillator \cite{HuertaAlderete2021p,HuertaAlderete2018p11572}. For the digital outcome, we measure all qubits in the $z$--basis, for each time $t$ with 5000 measurements per circuit.
The para-Fermi number is identified for each experimental shot according to the qubit mapping.
Once we measure $\langle \hat{\sigma}_{z}\rangle$ in all qubits it is easy to calculate $\langle \hat{\mathcal{N}}_{pF} \rangle$.
\begin{figure}[h]
	\includegraphics[scale=1]{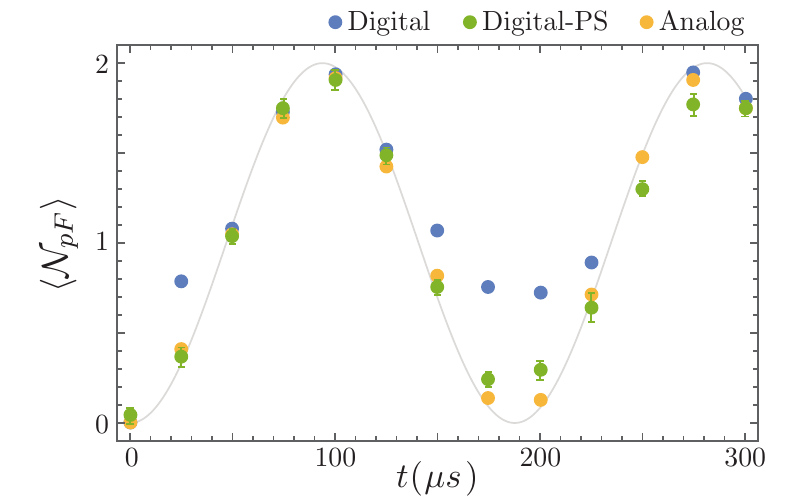}
	\caption{Evolution of the para-Fermi number operator $\langle \hat{\mathcal{N}}_{pF} \rangle$ of order 2. The digital outcome after SPAM correction (blue) and post-selection (green) is compared to the recent analog simulation (yellow) \cite{HuertaAlderete2021p}. Continuous line corresponds to the numerical simulation of the system. The error bars show statistical $1\sigma$ uncertainties.}
	\label{fig:Fig4}
\end{figure}
Figure \ref{fig:Fig4} shows the comparison of the evolution of the para-Fermi number operator $\langle \hat{\mathcal{N}}_{pF}\rangle$ of order 2 in the main two flavors for quantum simulation: the outcome of the digital simulation before (blue) and after (green) post-selection, and analog simulation (yellow) from Ref. \cite{HuertaAlderete2021p}. 
The continuous line corresponds to a classical numerical simulation of the system.
The evolution of the para-Fermi number operator for the lowest energy state produces a single frequency oscillation of the number operator.
For an increased order (not shown here), a multiple-frequency beat appears \cite{HuertaAlderete2018p11572,HuertaAlderete2021p}. 
The overall performance of the digital simulation after post--selection and the analog compare well for this particular case.
However, given the fact that the number of gates is 9 for this example and grows polynomially with the number of qubits involved, i. e. the dimension of the para-Fermi Hilbert space, experimental errors increase with increased para-Fermi order.

\begin{figure}[t]
	\includegraphics[scale=1]{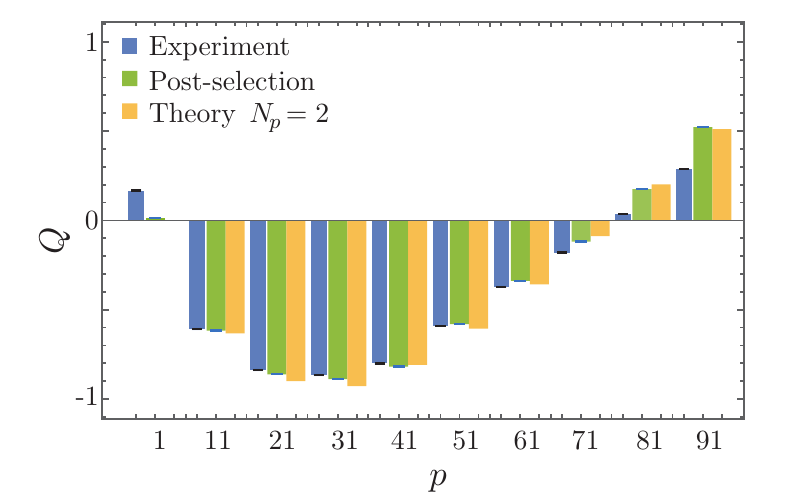}
	\caption{Mandel $Q$ parameter of a para-Bose state with complex parameter $\alpha=0.3$ and order $p$. Yellow bars correspond to the simulation of the three-qubit system and are compared to the experimental results after SPAM correction (blue) and post-selected data (green). The error bars show statistical $1\sigma$ uncertainties.}
	\label{fig:Fig3}
\end{figure}

The same decomposition can be used to simulate para-Bose oscillators of order $p$ by choosing adequate angles for the single-- and two--qubit operations.
We present the experimental results from a three--qubit system ($N_{p}=2$) with coherent parameter $\alpha=0.3$ and order $p$.
The system is prepared in the vacuum para-Bose Fock state, $\vert p; 0 \rangle$, as in Fig. \ref{fig:fig1}, followed by circuit that describes a para-Bose displacement operator for $N_{p}=2$, see Fig \ref{fig:Fig3}.
We measure all qubits in the $z$--basis, for each value of $p$ with 5000 measurements per circuit.
The para-Bose number is identified for each experimental shot according to the qubit mapping.
Once we measure $\langle \hat{\sigma}_{z}\rangle$ in all qubits it is easy to calculate $\langle \hat{\mathcal{N}}_{pB} \rangle$ and $\langle \hat{\mathcal{N}}_{pB}^2\rangle$ to obtain the Mandel Q parameter. 
Figure \ref{fig:Fig3} shows the comparison between the Mandel Q parameter calculated by using Eq. (\ref{eq:MandelQ}) and the experimental result obtained with the quantum circuit that describes the displacement operator, Eq. (\ref{eq:displacement}), and post-selected data. Even though the size of the Hilbert space is quite small, the experiment reproduces the transition from sub-- to super--Poissonian statistics, characteristic of a para-Bose system, very well. Furthermore, the results, after post-selection, provides the expected result for a coherent state, this is a Poissonian number distribution with complex parameter $\alpha$ and $p=1$.

In general, as we increase the number of qubits involved in our mapping, the number of operators needed to decompose the displacement operator increases as $N_{p}(N_{p}+1)$. These operators include interactions from two- to ($N_{p}+1$)-qubits.
Each interaction can be related to a multi-qubit gate. The multi-qubit gates follow the structure $(XZ\cdots ZX +YZ\cdots ZY)$ for even and $(XZ\cdots ZY +YZ\cdots ZX)$ for odd numbers of qubits involved.

In order to be able to scale this simulation, we now develop an optimal way to describe a gate decomposition for a multi-qubit operation when we increase the dimension of our Hilbert space.

Consider the next para-Fermi order, $p=4$, which needs 5 qubits. The commutation relations for the extended system are given in the Table \ref{tab:tab1}.
For larger systems, we express the interactions where two ($\hat{u}_{n}$), three ($\hat{v}_{n}$), $\dots$ $N_{p}+1$ consecutive neighbors are involved, as multi-qubit operations given by:
\begin{eqnarray}
\hat{u}_{n} &=& \sigma_{x}^{n} \sigma_{x}^{n+1} + \sigma_{y}^{n} \sigma_{y}^{n+1}, \nonumber\\
\hat{v}_{n} &=& \sigma_{x}^{n} \sigma_{z}^{n+1} \sigma_{y}^{n+2} - \sigma_{y}^{n} \sigma_{z}^{n+1} \sigma_{x}^{n+2},\nonumber\\
\hat{w}_{n} &=& \sigma_{x}^{n} \sigma_{z}^{n+1} \sigma_{z}^{n+2} \sigma_{x}^{n+3} + \sigma_{y}^{n} \sigma_{z}^{n+1} \sigma_{z}^{n+2} \sigma_{y}^{n+3},\nonumber\\
\hat{a}_{n} &=& \sigma_{x}^{n} \sigma_{z}^{n+1} \sigma_{z}^{n+2} \sigma_{z}^{n+3} \sigma_{y}^{n+4} + \sigma_{y}^{n} \sigma_{z}^{n+1} \sigma_{z}^{n+2} \sigma_{z}^{n+3} \sigma_{x}^{n+4}, \nonumber \\
& \vdots & 
\end{eqnarray}
From here, it is easy to identify operators $\left\{ \hat{\mathcal{A}}, \hat{\mathcal{B}}, \hat{\mathcal{C}}\right\}$ that satisfy the Jacobi identity, and hence use this set of operators to give the exact decomposition of the displacement operator, $\hat{D}_{\vartheta}$.
As before, we can use this set of operators to factorize the displacement operator.
All steps involved in the simulation (i.e. the preparation of the initial state, evolution, measurement, and measurement control) can be performed with a polynomial number of gates.
\newpage
\begin{widetext}
\begin{center}
\setlength{\tabcolsep}{8pt}
\renewcommand{\arraystretch}{1.5}
\begin{table}[h]
	\caption{Table of commutators for a five-qubit para-particle representation.}
	\begin{center}
		\begin{tabular}{ c|cccccccccc }
			$\left[ \cdot , \cdot\right]$ & $\hat{u}_{0}$ & $\hat{u}_1$ & $\hat{v}_{0}$ & $\hat{u}_{2}$ & $\hat{v}_{1}$ & $\hat{w}_{0}$ & $\hat{u}_{3}$ & $\hat{v}_{2}$ & $\hat{w}_{1}$ & $\hat{a}_{0}$ \\ \hline
			$\hat{u}_{0}$ & 0 & $2 i \hat{v}_{0}$ & $-2i\hat{u}_{1}$ & 0 & $-2i\hat{w}_{0}$ & $2i\hat{v}_{1}$ & 0 & 0 & $2 i \hat{a}_{0}$ & $-2i \hat{w}_{1}$\\ 
			$\hat{u}_{1}$ && 0 & $2i\hat{u}_{0}$ & $2i\hat{v}_{1}$ & $-2i\hat{u}_{2}$ & 0 & 0 & $-2i\hat{w}_{1}$ & $2i\hat{v}_{2}$ & 0\\
			$\hat{v}_{0}$ &&& 0 & $-2i\hat{w}_{0}$ & 0 & $2i\hat{u}_{2}$ & 0 & $-2i\hat{a}_{0}$ & 0 & $2i\hat{v}_{2}$\\ 
			$\hat{u}_{2}$ &&&& 0 & $2i\hat{u}_{1}$ & $-2i\hat{v}_{0}$ & $2i\hat{v}_{2}$ & $-2i\hat{u}_{3}$ & 0 & 0 \\
			$\hat{v}_{1}$ &&&&& 0 & $-2i\hat{u}_{0}$ & $- 2i\hat{w}_{1}$ & 0 & $2i\hat{u}_{3}$ & 0 \\ 
			$\hat{w}_{0}$ &&&&&& 0 & $2i\hat{a}_{0}$ & 0 & 0 & $- 2i\hat{u}_{3}$\\ 
			$\hat{u}_{3}$ &&&&&&& 0 & $2i\hat{u}_{2}$ & $-2i\hat{v}_{1}$ & $2i\hat{w}_{0}$\\ 
			$\hat{v}_{2}$ &&&&&&&& 0 & $-2i\hat{u}_{1}$ & $-2i\hat{v}_{0}$\\ 
			$\hat{w}_{1}$ &&&&&&&&& 0 & $2i\hat{u}_{0}$\\ 
			$\hat{a}_{0}$ &&&&&&&&&& 0 \\ 
		\end{tabular}
	\end{center}
	\label{tab:tab1}
\end{table}
\end{center}
\end{widetext}

\section{Conclusion}
We realize the digital quantum simulation of para-particle oscillators.
We discuss the steps for mapping, choosing a particular set of parameters and implementing their circuit-based representation on a trapped ion quantum computer.
For finite representations, as is the case for para-fermions, it is easy to decompose the propagator into native operations, however, due to the fast scaling of the number of gates with the system size, an analog simulation of the system might be a better alternative than the digital one. 
Even though the present comparison corresponds to the smallest case, we already see the strength of each simulation.
For the circuit-based implementation, the circuit depth does not change for the different evolution times, which means errors in the measurement will be related only to inherent errors in the gates, and strategies to mitigate them already exist  \cite{Czarnik2021p592,Bultrini2021p}.
Consequently, longer evolution times can be reached without being limited by coherence as in the analog simulation.
However, when increasing the para-Fermi order and, as a consequence, the system size, cumulative errors will have a bigger impact as the number of gates grows polynomially.
Here, an analog simulation will be advantageous.
In the para-Bose case, the measured statistical properties compare well with the theoretical prediction.
Here, the measurement of interest sets the cut-off of the Hilbert space and, therefore, the resources needed for its implementation.
This kind of measurements haven't been implemented on an analog simulation because of their experimental complexity.
Also, the digital implementation has the advantage of not being limited to para-Bose particles of only even order as in the analog experimental realization reported to date.
Our results light the way to the realization of larger para-particle systems as more advanced quantum computing hardware becomes available.

\begin{acknowledgments}
	This work received support from the National Science Foundation via the Physics Frontier Center at the Joint Quantum Institute (PHY-1430094), and through the Quantum Leap Challenge Institute for Robust Quantum Simulation (OMA-2120757), from the Maryland-ARL Quantum Partnership (W911NF1920181), and from the DOE Office of Science, Office of Nuclear Physics (DE-SC0021143). A. M. G. is supported in part by a Joint Quantum Institute Postdoctoral Fellowship.
\end{acknowledgments}

\nocite{*}
\bibliography{references}

\end{document}